\documentclass[twocolumn,superscriptaddress,showpacs,preprintnumbers,amsmath,amssymb,prl]{revtex4}
\usepackage{graphicx}
\usepackage{dcolumn,xcolor,ulem}
\usepackage{amsmath} 
\usepackage{amssymb}
\usepackage{amsfonts}
\usepackage{bm}
\usepackage[latin1]{inputenc}

\newcommand\gp{\dot\gamma}

\newcommand\Rey{\mbox{\textit{Re}}}  % Reynolds number
\newcommand\Wei{\mbox{\textit{Wi}}}  % Weissenberg number
\newcommand\Ta{\mbox{\textit{Ta}}}  % Taylor number
\newcommand\El{\mathcal{E}} %Elasticity number

\begin{document}

%\widowpenalty=1000
%\clubpenalty=1000
%\preprint{APS/123-QED}

\title{Shear-induced structures versus flow instabilities}

\author{M.A.~Fardin}
\email{marcantoine.fardin@ens-lyon.fr}
\affiliation{Universit\'e de Lyon, Laboratoire de Physique, \'Ecole Normale Sup\'erieure de Lyon, CNRS UMR 5672, 46 All\'ee d'Italie, 69364 Lyon cedex 07, France}
\affiliation{The Academy of Bradylogists}
\author{C.~Perge}
\affiliation{Universit\'e de Lyon, Laboratoire de Physique, \'Ecole Normale Sup\'erieure de Lyon, CNRS UMR 5672, 46 All\'ee d'Italie, 69364 Lyon cedex 07, France}
\author{N.~Taberlet}
\affiliation{Universit\'e de Lyon, Laboratoire de Physique, \'Ecole Normale Sup\'erieure de Lyon, CNRS UMR 5672, 46 All\'ee d'Italie, 69364 Lyon cedex 07, France}
\author{S.~Manneville}
\affiliation{Universit\'e de Lyon, Laboratoire de Physique, \'Ecole Normale Sup\'erieure de Lyon, CNRS UMR 5672, 46 All\'ee d'Italie, 69364 Lyon cedex 07, France}
%\affiliation{Institut Universitaire de France}

\date{\today}

\begin{abstract}
The Taylor-Couette flow of a dilute micellar system known to generate shear-induced structures is investigated through simultaneous rheometry and ultrasonic imaging. We show that flow instabilities must be taken into account since both the Reynolds number and the Weissenberg number may be large. Before nucleation of shear-induced structures, the flow can be inertially unstable, but once shear-induced structures are nucleated the kinematics of the flow become chaotic, in a pattern reminiscent of the inertio-elastic turbulence known in dilute polymer solutions. We outline a general framework for the interplay between flow instabilities and flow-induced structures.
\end{abstract}

\pacs{83.80.Qr, 47.20.-k, 47.27.-i, 47.50.Gj }
\maketitle

Beside their tremendous industrial importance in applications such as detergence, oil recovery, or drag reduction, micellar systems have long been used as a model system for rheological research~\cite{Reh91}. In particular some surfactant systems are known to form rodlike micelles, which can grow to become wormlike and entangled when the concentration in surfactant and/or salt increases, or when the temperature decreases~\cite{Ber2006,Cat2006}. The dilute regime of rodlike micelles has been studied in the context of highly elastic shear-induced structures (SIS) and associated shear-thickening, whereas the so-called ``semi-dilute'' and ``concentrated'' regimes have been studied in the context of shear-banding, which is an extreme form of shear-thinning where the velocity gradient field becomes inhomogeneous even in simple shear~\cite{Ler2010}. 

On the one hand, shear-banding micellar systems have a high viscosity, such that inertial flow instabilities can generally be neglected. Nevertheless semi-dilute and concentrated solutions have high normal stresses, such that the Weissenberg number is large and purely elastic instabilities can develop~\cite{Fie2010,Far2012d,Nic2012}, similar to the case of polymeric Boger fluids~\cite{Lar90,Lar92,Mor2007}. On the other hand, the possibility for flow instabilities has never been considered thoroughly in dilute, shear-thickening systems even though these systems have both large Reynolds ($\Rey$) and Weissenberg ($\Wei$) numbers~\cite{Ler2010}. The aim of the present Letter is to fill this gap by showing that both inertial and elastic instabilities are constantly encountered in the Taylor-Couette flow of a well-known dilute micellar system.

\begin{figure}
\centering
\includegraphics[width=7cm,clip]{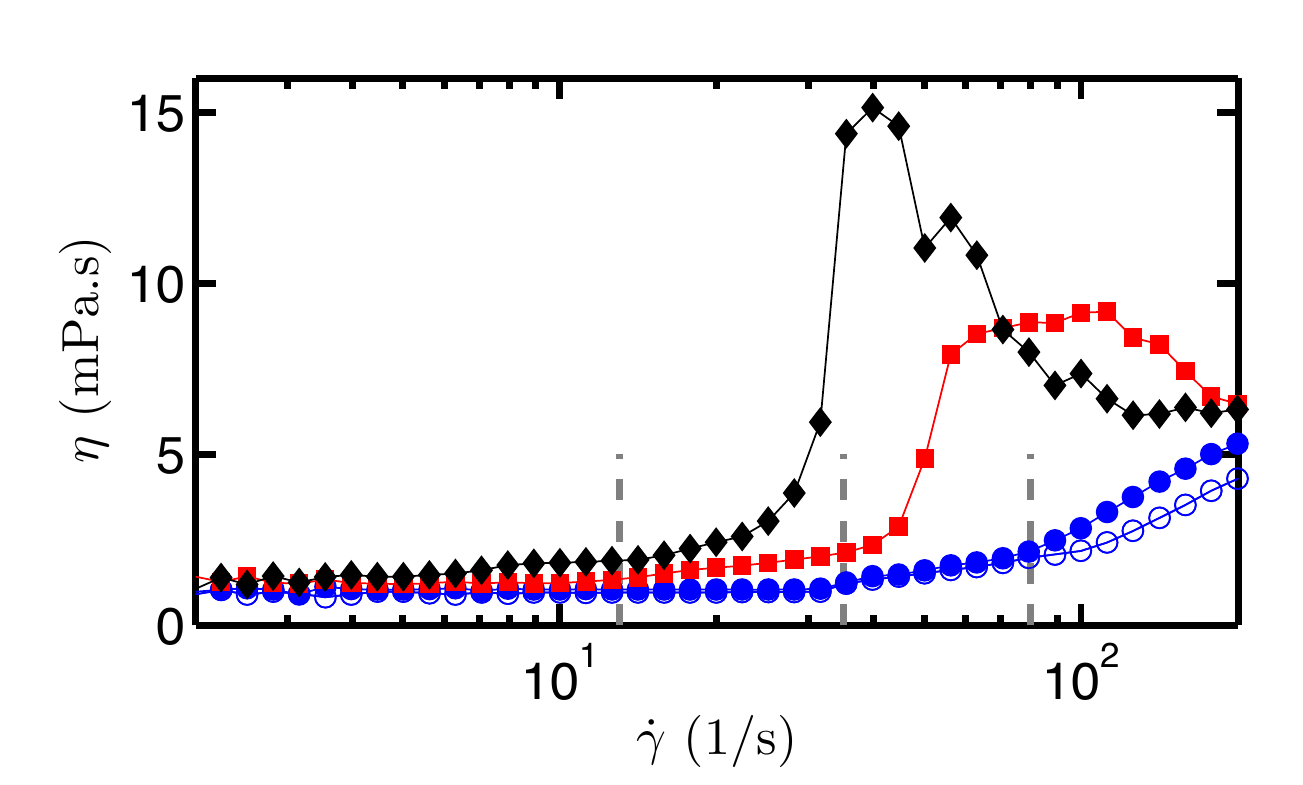}
\caption{Viscosity $\eta$ vs shear rate $\gp$ at $T=20$, 25, and 30$^\circ$C (filled symbols, from top left to bottom right) for CTAT at 0.16\%~wt. seeded with 1\%~wt. polyamide spheres (see text). Open symbols correspond to a CTAT sample free of contrast agents at $T=30^\circ$C. The addition of polyamide spheres increases the viscosity by about 15\% but does not change the overall behavior. Data points correspond to averages over the last 5~s of shear rate steps of total duration 15~s. Vertical dashed lines indicate the onset of shear-thickening $\gp_c\simeq 13$, 35, and 80~s$^{-1}$ respectively inferred at $T=20$, 25, and 30$^\circ$C from start-up experiments of typical duration 500~s .
\label{fig1}}
\end{figure} 

The sample under study is made of 0.16\%~wt. hexadecyltrimethylammonium p-toluenesulfonate (CTAT) in water. For this system, which has become a benchmark example of dilute micellar fluids~\cite{Ler2010}, the overlap concentration is estimated at $c^\star \simeq 0.5$~wt.~\% and shear-thickening is found over the range 0.05-0.8\%~wt.~\cite{Gam99,Tru2000}. Figure~\ref{fig1} shows the viscosity $\eta$ of our sample as a function of the applied shear rate $\gp$ for three different temperatures $T$, exhibiting the typical behavior of shear-thickening, dilute surfactant systems~\cite{Ler2010}: a zero-shear viscosity close to the viscosity of water; then a jump in $\eta$ at a characteristic shear rate $\gp_c$ that increases with $T$; finally, a shear-thinning viscosity branch at high shear rates. This behavior was first explained by postulating the formation of SIS~\cite{Reh82}: above $\gp_c$, micelles grow in length and undergo a transition from rodlike to wormlike aggregates. This microscopic scenario was confirmed through neutron and light scattering experiments~\cite{Ler2010}. The shear-induced state can then be shear-thinning due to the increasing alignment of the worms.

In the present experiments the fluid is sheared in a Taylor-Couette (TC) device adapted to a rheometer (ARG2, TA Instruments) and with dimensions (height $H=60$~mm, gap $d=2$~mm, radius of the inner rotating cylinder $R_i=23$~mm) that ensure the ``small gap approximation'', $\Lambda\equiv d/R_i \simeq 0.087 \ll 1$, without any strong end effect at the top and bottom boundaries ($d/H\ll 1$), so that the laminar base flow is a simple shear with $\gp=U/d=\Omega R_i/d$, where $U$ and $\Omega$ are the rotor linear and angular velocities respectively. We recall that in a TC device with inner rotation, Taylor showed that a Newtonian fluid becomes unstable and develops a secondary flow made of toroidal counter-rotating vortices~\cite{Tay23}. Larson, Shaqfeh and Muller discovered that non-Newtonian fluids can develop a similar vortex flow not driven by inertia, but driven by elasticity~\cite{Lar90}. In both cases the instability develops when the Taylor number $\Ta$ exceeds a given threshold $\Ta_c$. In the purely inertial case $\Ta = \Lambda^{1/2} \Rey$, with $\Rey\equiv\tau_1 \gp = (\rho d^2/\eta) (U/d) = \rho d U/\eta$ and $\Ta_c\simeq 41$~\cite{Tay23,Don60}, while in the purely elastic case $\Ta=\Lambda^{1/2} \Wei$, with $\Wei\equiv\tau_2 \gp$, where $\tau_2$ is a characteristic polymeric relaxation time~\cite{Lar90,McK96} instead of the viscous dissipation time $\tau_1$, and $\rho\simeq 10^3$~kg.m$^{-3}$ is the fluid density. In the latter case, the value of $\Ta_c$ depends on the constitutive relation of the non-Newtonian fluid, e.g., linear stability theory predicts $\Ta_c\simeq 6$ for the Upper-Convected Maxwell model~\cite{Lar90}. More generally the balance between elasticity and inertia can be estimated by the elasticity number $\El \equiv \Wei/\Rey = \tau_2/\tau_1$. Although predicting the onset of instability for arbitrary elasticity is not yet possible, a range of inertio-elastic instabilities is expected but remains mostly unexplored~\cite{Mul2008}.

Since we expect secondary flows to emerge, our geometry is equipped with a recently developed two-dimensional ultrasonic velocimetry technique that allows the simultaneous measurement of 128 velocity profiles over 30~mm along the vertical direction in the TC geometry~\cite{Gal2013}. We use ultrafast plane wave imaging and cross-correlation of successive images~\cite{San2001} to infer velocity maps from the echoes backscattered by acoustic contrast agents seeding the fluid, namely 1\%~wt. polyamide spheres (Arkema Orgasol 2002 ES 3 Nat 3, mean diameter 30~$\mu$m, relative density 1.03), which do not affect significantly the rheological behavior of our solution (see Fig.~\ref{fig1}). This technique yields the component $v_y(r,z)$ of the velocity vector, $\mathbf{v}=(v_r,v_\theta,v_z)$ in cylindrical coordinates, projected along the acoustic propagation axis $y$ as a function of the radial distance $r$ to the rotor and of the vertical position $z$ with a temporal resolution down to 50~$\mu$s~\cite{Gal2013}. The acoustic axis $y$ is horizontal and makes an angle $\phi\simeq 10^\circ$ with the normal to the outer cylinder so that $v_y=\cos\phi\, v_r + \sin\phi\, v_\theta$. Finally we define the measured velocity as $v=\frac{v_y}{\sin\phi}=v_\theta+\frac{v_r}{\tan\phi}$, which coincides with the azimuthal velocity $v_\theta$ in the case of a purely azimuthal flow $\mathbf{v}=(0,v_\theta,0)$. More generally $v$ combines contributions from both azimuthal and radial velocity components. Nevertheless, close to instability onset, secondary flows are usually much weaker than the main flow, such that $v\simeq v_\theta$, as checked recently on an inertially unstable Newtonian fluid and on an elastically unstable shear-banding fluid~\cite{Gal2013,Per2013}.

\begin{figure}
\centering
\includegraphics[width=6.8cm,clip]{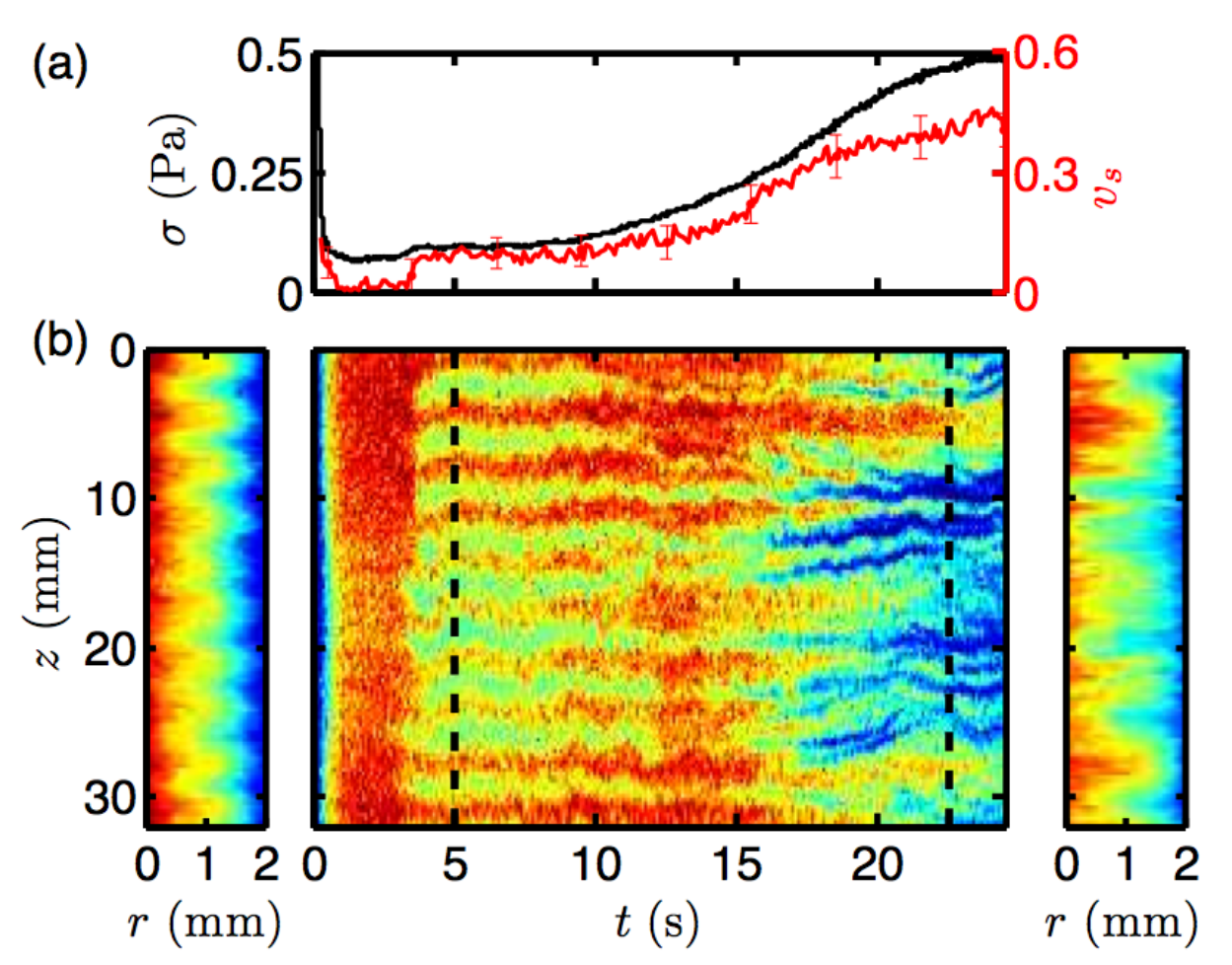}
\caption{Spatiotemporal dynamics of CTAT at $T=25^\circ$C for $\gp=50$~s$^{-1}$ (see also Supp. Movie~1). The flow is first inertially unstable and then inertio-elastically unstable due to SIS formation. (a)~Global shear stress $\sigma(t)$ measured by the rheometer (in black) and dimensionless slip velocity $v_s(t)$ (in red)~\cite{Analysis}. (b) Spatiotemporal diagram (center) of the velocity $v(r_0,z,t)$ at $r_0=0.20$~mm from the rotor. The dotted lines show the times $t_0=5$ and 22.5~s corresponding to the velocity maps $v(r,z,t_0)$ shown on the left and right respectively. The color scale is linear and goes from 0 to $\Omega R_i$ for the velocity maps and from $0.5 \Omega R_i$ to $\Omega R_i$ for the diagram. The vertical axis $z$ is oriented downwards with $z=0$ being taken at about 6 mm from the top of the TC cell.
\label{fig2}}
\end{figure} 

Figure~\ref{fig2} reports the start-up flow of CTAT at $T=25^\circ$C for $\gp=50$~s$^{-1}$ (see also Supp. Movie~1). At very short times, a laminar boundary layer extends from the inner cylinder to the outer cylinder. A Taylor vortex flow (TVF) then develops for $t\gtrsim 3$~s, deforming the main flow which becomes periodic along $z$: slow moving fluid is brought inward in regions of centripetal radial flow and fast moving fluid is pushed outward in regions of centrifugal radial flow. This initial sequence of events would be exactly similar if the fluid was pure water~\cite{Ako2003,Gal2013}. Even though such a TVF had never been reported before for dilute micelles, it should not be surprising since, assuming the fluid to be influenced only by inertia in this initial sequence, we have $\Ta=\Lambda^{1/2}\Rey \simeq 60 > \Ta_c$ for $\gp=50$~s$^{-1}$. In computing $\tau_1$, we have used the dynamic viscosity relevant to the short-time behavior, i.e. the zero-shear viscosity $\eta_0\simeq 1$~mPa.s. As shown in Fig.~\ref{fig2}(a) the onset of TVF at $t\simeq 3$~s corresponds to a slight increase of the shear stress $\sigma$ (after an initial spike due to the feedback with the rheometer inertia). This first stress increase is simply due to the formation of vortices breaking the viscometric assumption~\cite{Gal2013}. In contrast, for $t\gtrsim 10$~s, the stress (or alternatively the viscosity) climbs up much more dramatically. Since $\gp =50$~s$^{-1}$ falls into the shear-thickening range for $T=25^\circ$C (see Fig.~\ref{fig1}), this next sequence of events can clearly be attributed to slow SIS formation. Meanwhile the structure of the vortex flow is disrupted. The formerly well-defined wavelength and amplitude of the main flow shown in Fig.~\ref{fig2}(b,left) are lost and the flow becomes chaotic-like in Fig.~\ref{fig2}(b,right). This latter state is reminiscent of the inertio-elastic turbulent state called ``disordered oscillations''~\cite{Gro96} or ``elastically dominated turbulence''~\cite{Dut2013}. While iso-velocity lines in the initial state can be approximated by harmonic functions, the new secondary flows associated with the SIS deform the main flow intermittently. The state in Fig.~\ref{fig2}(b,right) is representative of the asymptotic turbulent flow. It continuously generates large fluctuations in the viscosity and stress, which have been reported before but never accounted for in terms of elastic turbulence~\cite{Ler2010}. Note also that the turbulent nature of the flow can locally and transiently generate plug flow profiles that may explain some earlier 1D velocity measurements~\cite{Koc98,Hu98}. At lower shear rates, e.g. $\gp=20$~s$^{-1}<\gp_c$ and $\Ta\simeq 24<41$, the flow is below both thresholds for TVF and SIS formation and remains purely azimuthal as shown in Supp. Fig.~1. 

\begin{figure}
\centering
\includegraphics[width=6.8cm,clip]{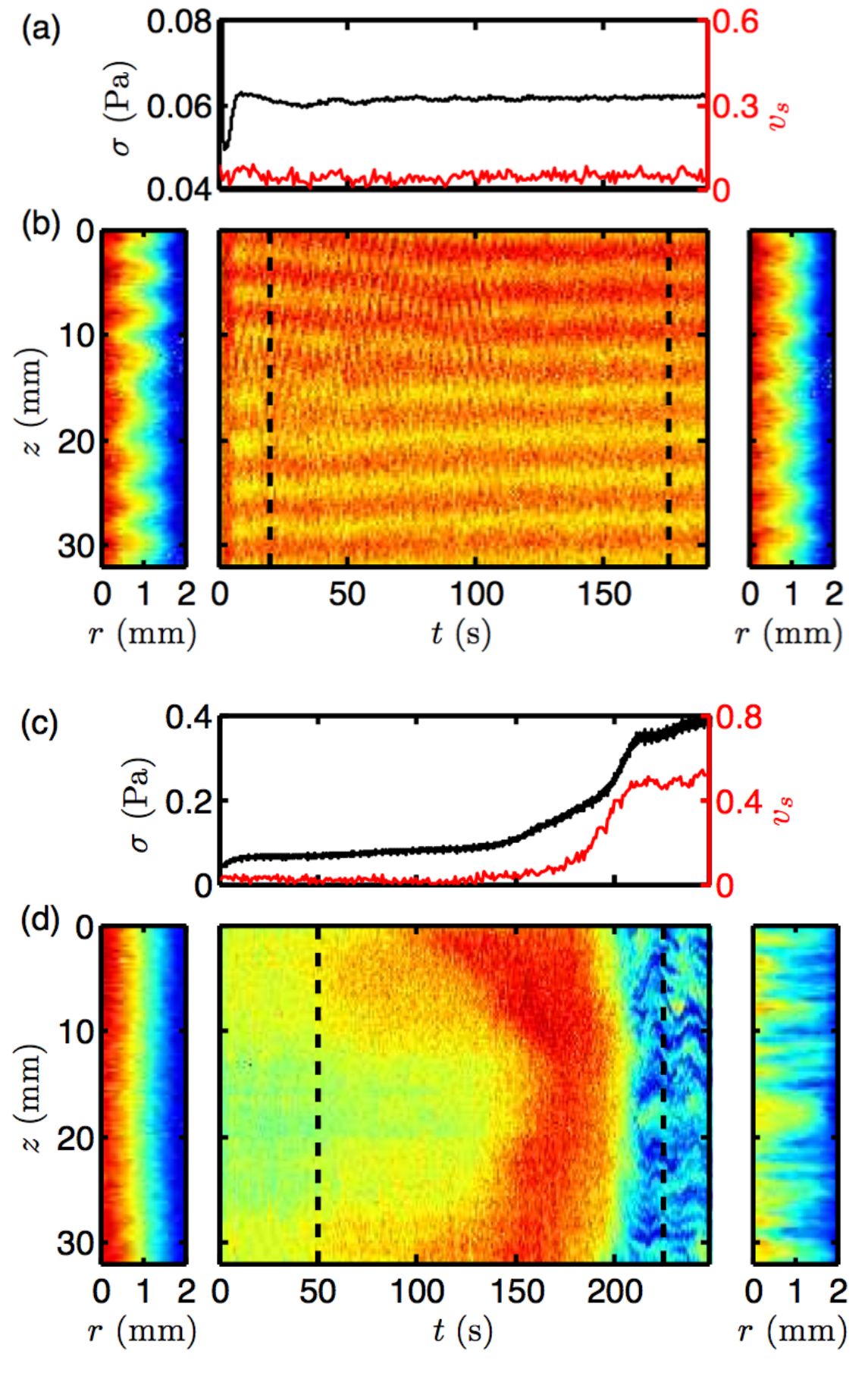}
\caption{Same as Fig.~\ref{fig2} for (a,b) $T=30^\circ$C and $\gp=40$~s$^{-1}$ showing only inertial instability and (c,d) $T=20^\circ$C and $\gp=20$~s$^{-1}$ where the flow slowly develops SIS and the associated inertio-elastic turbulence without initial TVF (see also Supp. Movies~2 and 3 respectively). In (b) $r_0=0.40$~mm and $t_0=20$ (left) and 175~s (right). In (d) $r_0=0.53$~mm and $t_0=50$ (left) and 225~s (right). The color scale goes from 0 to $\Omega R_i$ for all velocity maps and from $\Omega R_i/8$ (resp. $\Omega R_i/4$) to $\Omega R_i$ for the diagram in (b) [resp. (d)].
\label{fig3}}
\end{figure}

As it turned out for $T=25^\circ$C, the critical shear rate $\gp_c$ for SIS formation and the critical shear rate $\gp_{\rm TVF}\equiv \Ta_c/(\tau_1\Lambda^{1/2})$ for the onset of TVF are about the same value $\gp_c \simeq \gp_{\rm TVF} \simeq 35$~s$^{-1}$ in our TC geometry ($\Lambda^{1/2}\simeq 0.29$). In order to separate the inertial TVF and the turbulence associated with SIS more readily, we reproduced similar shear start-up protocols at two other temperatures, $T=20$ and $30^\circ$C, as shown in Fig.~\ref{fig3}. Increasing the temperature slightly lowers the zero-shear viscosity (see Fig.~\ref{fig1}) so that $\gp_{\rm TVF}$ only decreases from 45~s$^{-1}$ at $20^\circ$C to 33~s$^{-1}$ at $30^\circ$C. In contrast, the same temperature change has a much stronger impact on $\gp_c$. As reported extensively in the literature~\cite{Ler2010}, lowering the temperature leads to easier SIS formation hence shifting $\gp_c$ to lower values. Figure~\ref{fig1} indicates $\gp_c\simeq 13$ and 80~s$^{-1}$ at $20^\circ$C and $30^\circ$C respectively. Therefore at the highest temperature, we should be able to observe TVF without SIS, whereas SIS without TVF may be expected at the lowest temperature. This scenario is fully confirmed in Fig.~\ref{fig3}(a,b) and (c,d) where spatiotemporal dynamics are compared for shear rates such that $\gp_{\rm TVF}< \gp<\gp_c$ and $\gp_c< \gp<\gp_{\rm TVF}$ at $T=30$ and $20^\circ$C respectively.

The fact that SIS and TVF can occur separately is an indication that these two phenomena are not consequences of one another. SIS do not need TVF to nucleate, which suggests that the out-of-equilibrium growth of the worms is driven by the base shear flow, as usually postulated~\cite{Ler2010}. 
Early velocity measurements at a single height $z$ reported that SIS first form at the inner wall and generate significant slip on this wall~\cite{Koc98,Hu98,Bol97}. The dimensionless slip velocities $v_s$ \cite{Analysis} shown in Figs.~\ref{fig2}(a) and \ref{fig3}(c) confirm that the onset of wall slip is concomitant with SIS formation, whereas no noticeable wall slip is reported in the presence of TVF alone [see Fig.~\ref{fig3}(a)].

Of course TVF does not need SIS since it can occur even in simple molecular fluids. Moreover the spatiotemporal structure of the flow in Figs.~\ref{fig2}(b,right) and \ref{fig3}(d,right) are very similar so that TVF appears to have negligible impact on the asymptotic turbulent flow after SIS nucleation. It seems that before the nucleation of SIS inertia is dominating the instability of the flow, whereas once SIS are formed elasticity dominates. In a shear-thickening dilute surfactant solution, both $\tau_1$ and $\tau_2$ depend on $\gp$ and $t$ so that $\Rey$, $\Wei$, and $\El$ also depend on $\gp$ and $t$. To illustrate this, we evaluate a bulk-averaged value of $\El$ in the case of Fig.~\ref{fig2} ($T=25^\circ$C and $\gp=50$~s$^{-1}$). Assuming the fluid density $\rho$ to remain constant during SIS formation, we first estimate the Reynolds number by $\Rey \simeq \gp_t \rho d^2/\eta_t$, where $\gp_t(\gp,t)=\gp-v_s(t)\Omega R_i/d$ is the ``true'' shear rate corrected for wall slip and $\eta_t(\gp,t)=\sigma(t)/\gp_t(\gp,t)$ the corresponding ``true'' viscosity. This yields a decrease from $\Rey \simeq 200$ before SIS formation to $\Rey \simeq 7$ in the final state. Estimating the characteristic viscoelastic time $\tau_2$ of the SIS is more challenging. We take the longest time of the stress relaxation after flow cessation either before or after SIS formation, which yields $\Wei=\gp_t\tau_2\simeq  0$ before and $\Wei\simeq 100$ after SIS formation in accordance with the literature~\cite{Ler2010}. Thus in the early stages of the dynamics $\El\simeq 0$, i.e. inertia dominates, whereas after SIS formation $\El\gtrsim 10$, validating the dominance of elasticity. In some sense, the dynamics of Fig.~\ref{fig2}(b) can be seen as the superposition of the dynamics of Fig.~\ref{fig3}(b) and (d)~\cite{Remark}.

\begin{figure} 
\centering
\includegraphics[width=8cm,clip]{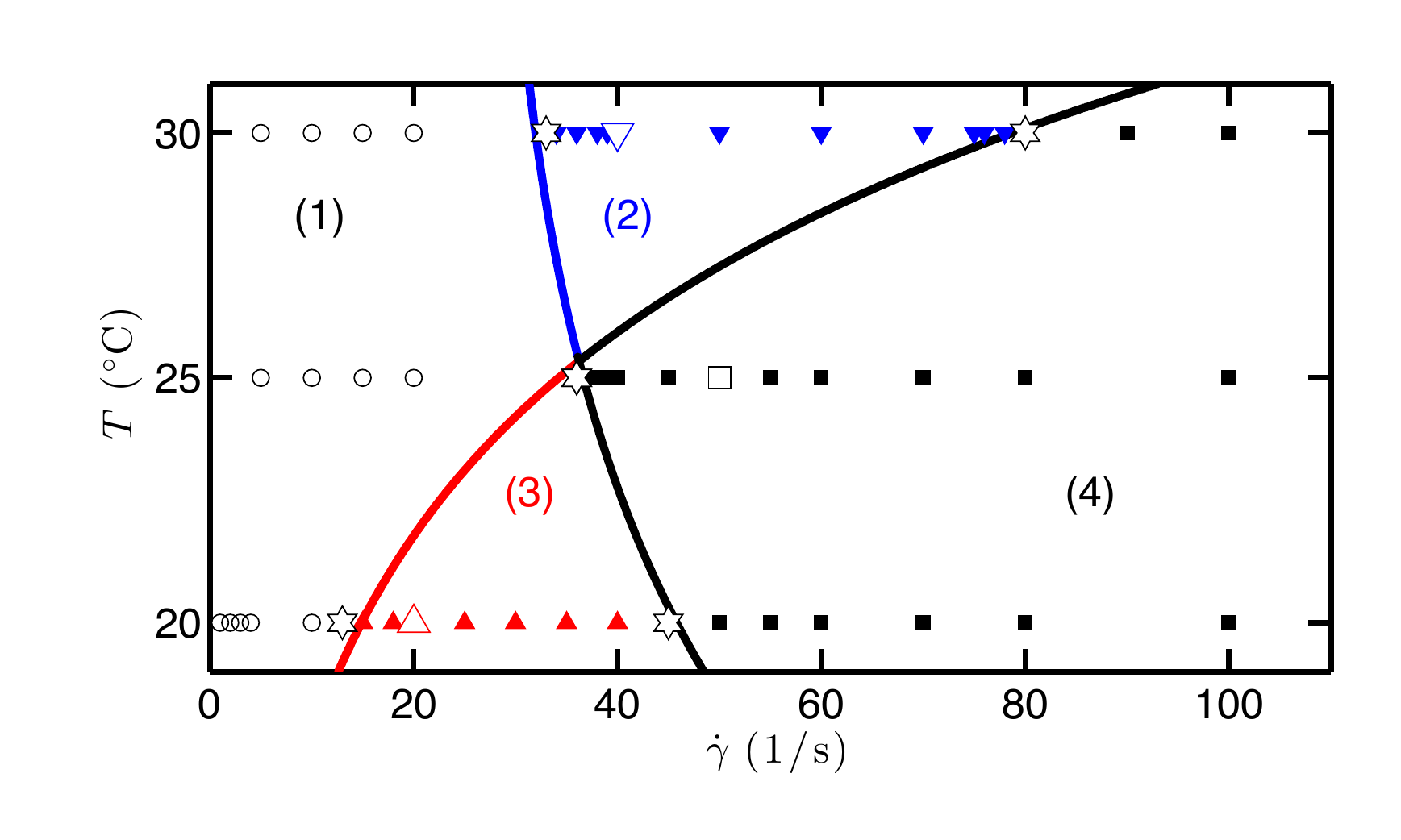}
\caption{Stability diagram of CTAT. (1)~Laminar flow ($\circ$). (2)~Inertial instability (TVF) without SIS ($\blacktriangledown$). (3)~Inertio-elastic instability of the SIS without initial TVF ($\blacktriangle$). (4)~Inertio-elastic instability of the SIS after initial TVF ($\blacksquare$). Larger empty symbols correspond to the experiments in Figs.~\ref{fig2} and \ref{fig3}. Stars show the transitions between the various regimes. Solid lines are guides to the eye.
\label{fig4}}
\end{figure} 

The stability diagram of Fig.~\ref{fig4} summarizes the interplay between inertial instability (TVF) and elastic instability of the SIS based on an extensive data set. We believe that such a diagram should be systematically sought for in complex fluids subject to flow instabilities in order to sort out the influences of flow-induced microscopic and macroscopic phenomena. Here we find that SIS are always unstable and systematically lead to elastic-like turbulence. Whether or not this is a feature common to all elastic SIS stands out as an open issue. Another fundamental question is the possibility to predict the various boundaries in Fig.~\ref{fig4} using phenomenological approaches or even microscopic models. Deriving an instability criterion based on the elasticity number $\El$ and accounting for both $\Rey$ and $\Wei$ thus appears as a first crucial theoretical step.

%\textbf{\normalsize{Acknowledgments }} \\
\begin{acknowledgments}
M.-A.F. and C.P. contributed equally to this work. The authors thank S. Lerouge for providing the CTAT sample and for motivating this study. This work was funded by the Institut Universitaire de France and by the European Research Council under the European Union's Seventh Framework Programme (FP7/2007-2013) / ERC grant agreement No.~258803. 
\end{acknowledgments}

\clearpage
\newpage
\setcounter{figure}{0}

\section{Supplemental material}
\begin{center}
{\bf Shear-induced structures versus flow instabilities}
\end{center}

\subsection*{ } 
Sup. Movies 1, 2 and 3 corresponding respectively to Fig.~2(a-b), Fig.~3(a-b) and Fig.~3(c-d) of the paper are available on request from the corresponding author. See the captions of the corresponding figures for details. The movies also display some velocity profiles in the bottom left quadrant. The square symbols correspond to examples of velocity profiles obtained at three different heights $z$, whereas the circles correspond to the velocity profile obtained by averaging along $z$. The line represents the purely azimuthal Couette flow profile expected for no-slip boundary conditions. 
\subsection*{ }

\begin{figure}[h]
\centering
\includegraphics[width=8cm,clip]{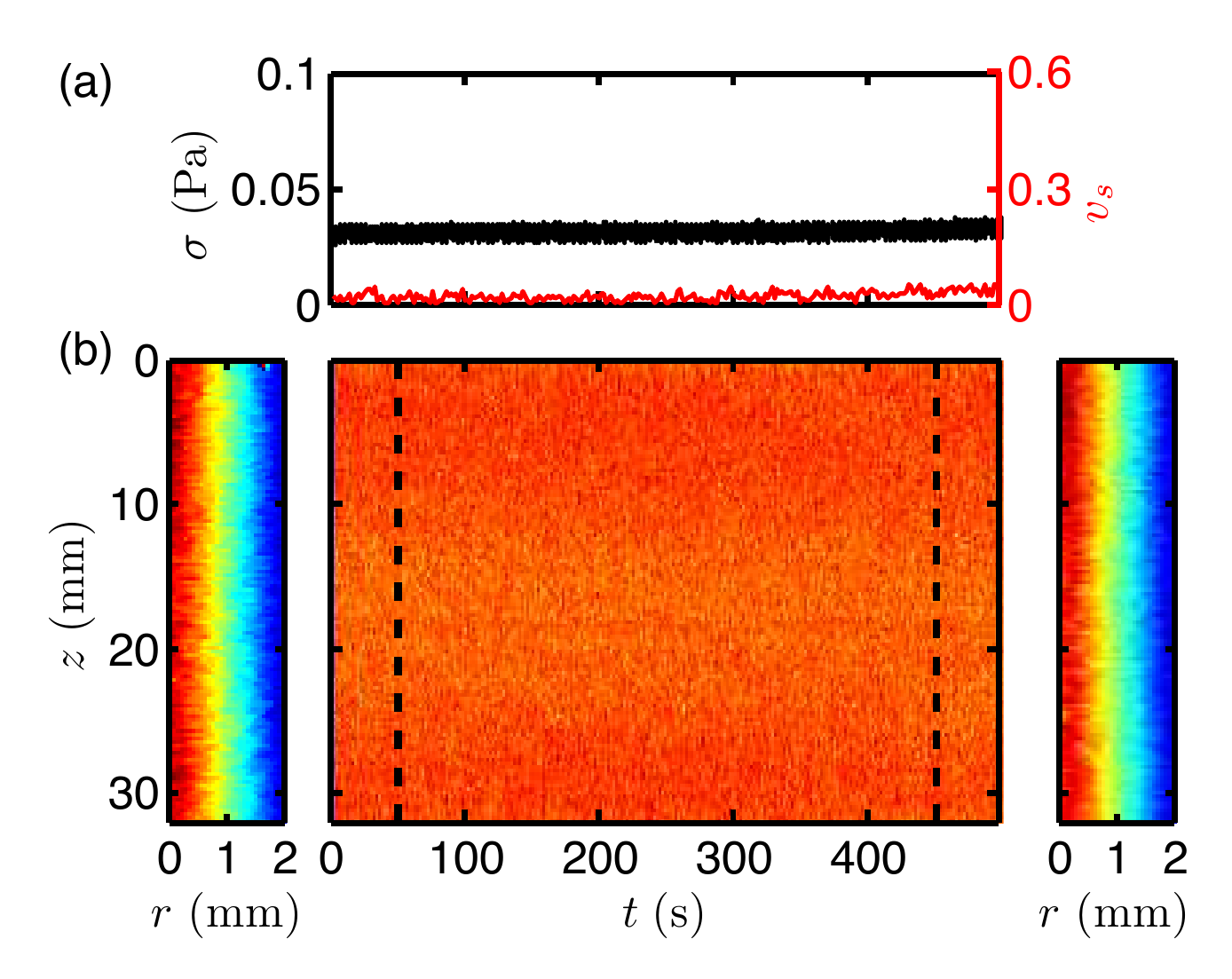}
\caption{Spatiotemporal dynamics of the flow of CTAT for T=$25^\circ$C, at $\gp=20$~s$^{-1}$, below the onsets of both TVF and SIS formation. The flow remains perfectly laminar with no noticeable wall slip. (a) Global shear stress $\sigma(t)$ measured by the rheometer (in black) and dimensionless slip velocity $v_s(t)$ (in red). (b) Spatiotemporal diagram (center) of the velocity $v(r_0,z,t)$ at $r_0=0.3$~mm from the rotor. The dotted lines show the times $t_0=50$ and $450$~s corresponding to the velocity maps $v(r,z,t_0)$ shown on the left and right respectively. The color scale is linear and goes from 0 to $\Omega R_i$.  
\label{supfig}}
\end{figure} 


\begin{thebibliography}{37}
\bibitem{Analysis} Linear fits of the velocity profiles $v(r,z)$ over about 300~$\mu$m in the radial direction $r$ are used to estimate the local shear rates close to each wall and extrapolate the fluid velocities $v(r=0,z)$ and $v(r=d,z)$. The dimensionless total slip velocity is then defined as $v_s=1 - [ \langle v(0,z)\rangle_z - \langle v(d,z)\rangle_z ] / \Omega R_i$ where the average is taken over the 128 simultaneous measurements along the vertical direction $z$. Error bars in Fig.~\ref{fig2}(a) show the standard deviation along $z$.
\end{thebibliography}
\end{document}